**Tunneling Devices with Perpendicular Magnetic Anisotropy Electrodes on Atomically Thin van der Waals Heterostructures**

H. Idzuchi[1], T. Taniguchi[2], K. Watanabe[2], and P. Kim[1*]

[1] *Department of Physics, Harvard University, Cambridge, MA 02138, USA*

[2] *National Institute for Materials Science, 1-1 Namiki, Tsukuba 305-0044, Japan*

[*]pkim@physics.harvard.edu

**Abstract**

We report the fabrication of perpendicular ferromagnetic electrodes for tunneling devices consist of van der Waals heterostructure. We found MgO/Co/Pt films on Hexagonal BN shows perpendicular magnetic anisotropy (PMA) with the easy axis perpendicular to the substrate. Vacuum annealing enhances the perpendicular anisotropy. The easy axis along the perpendicular direction persists up to room temperature with the Pt layer thickness ranges from 1.5 to 5 nm. We employed the PMA electrodes to construct tunneling devices on graphene and monolayer $WSe_2$, where spin injection characteristics and field effect transistor behavior were demonstrated without strong Schottky barrier formation.

Layered van der Waals (vdW) material provides unique opportunities ranging from unusual electronic, optical and spintronic properties [1]. For example, in semiconducting transition metal dichalcogenides (TMDs), such as $WSe_2$ and $MoS_2$, valley degree of freedom can be accessed by circularly polarized light and thus optoelectrical effect is widely studied [2,3]. In the monolayer (ML) limit, spin and valley degrees of freedom are coupled, which can be utilized for potential spintronics applications [4]. Most of the experimental reports on spin-valley coupled transport have been relied on polarized optical excitation [5]. Alternative access of spin-valley degrees of freedom can in principle realized by spin-polarized carriers injection. Since the valley degree of freedom is coupled to the perpendicular spin in ML TMD [2], to inject spin into the 2-dimensional (2D) channel, ferromagnet with perpendicular magnetic anisotropy is required. In this letter, we report highly efficient tunneling junctions fabricated between perpendicular magnetic film and atomically thin vdW materials including graphene and ML $WSe_2$ TMD semiconductor.

Perpendicular magnetic anisotropy (PMA), often induced by crystalline, interfacial and surface anisotropy has been studied long time [6-8]. A typical example of the crystalline perpendicular magnetic system is $L1_0$ ordered CoPt(001) and FePt(001) film or rare earth based material such as TbCoFe [9]. A widely used interface anisotropy system is Co/Pt or Co/Pd interface [7]. Strong perpendicular anisotropy can also be induced by Oxide/Fe or Oxide/Co system [10,11]. In order to combine PMA electrodes with 2D vdW channel materials, the following conditions should be considered: (i) compatible with the lift-off based fabrication process; (ii) formation of atomically sharp interfaces with vdW materials; (iii) low temperature (< 400 °C) fabrication process; and (iv) minimized oxidization and aging effect of the surface of the electrodes to form vdW interface with the channel. We identified MgO/Co/Pt films as an excellent candidate satisfying all the conditions above. Importantly, as we demonstrate in the following, the MgO/Co/Pt PMA layer can be fabricated on Hexagonal BN (hBN) surface, which has been served as a popular choice of dielectric and tunneling barrier in 2D materials [12], easily facilitating tunneling vdW heterostructures devices with PMA electrodes.

Samples were prepared on a Si substrate with a $SiO_2$ layer with the thickness of 285 nm. hBN flakes were mechanically exfoliated. The surfaces of the hBN samples were examined by atomic force microscope (AFM) before the deposition process of the PMA. Pt, Co and MgO layers were sputtered on $h$-BN flakes on the $SiO_2$/Si substrate. Pt and Co were sputtered with the DC power of 50 W (0.75 Å/s) and 100 W (0.43 Å/s). A MgO layer was sputtered with RF power of 120 W (0.050 Å/s) with a working pressure of 4 mTorr (Ar flow of 40 sccm) and a base pressure of $10^{-8}$ Torr. In order to test the magnetic properties of resulting film, we fabricate a mesoscopic sized (~ 10 $\mu m^2$) Hall bar device on the PMA film on a hBN flake. The inset of Figure 1(a) shows the device. A poly-methyl-methacrylate (PMMA) etch mask was patterned by electron beam lithography, and a rectangular shape was formed by Ar ion-milling. The temperature rise during ion-milling was kept to be minimized by controlling active etching time. We also avoid the $O_2$ plasma etching process to remove resist which may cause damage the surface of the device [13]

Figure 1(a) shows Hall signal of MgO/Co/Pt/hBN on $SiO_2$/Si with the thickness of MgO, Co and Pt, 1.5 nm, 1.0 nm and 5.0 nm respectively. The thickness of Co is fixed in this study. Hysteresis shows the perpendicular anisotropy of the film, exhibiting a clear signature of magnetization induced anomalous Hall effect. Importantly, annealing at high vacuum (~$10^{-7}$ Torr) at $T$ = 350 °C results in significant increase of the perpendicular anisotropy as shown in Fig. 1(b). We attribute this enhancement to the increase of the effective area of the Co-O bonding in our PMA layers [11,14]. Also, the better crystallite growth and more diffusion of Pt and Co atoms along the grain boundaries at higher temperatures [15] can contribute the enhanced PMA in the annealing process. It is notable that our measured perpendicular anisotropy of the MgO/Co/Pt perpendicular layer on the vdW hBN substrate is similar to those fabricated on a conventional substrate. Usually, MgO/Co/Pt perpendicular layer has been fabricated with a Ta metallic seed layer or a TaO$_x$ seed layer [16]. These films showed perpendicular easy axis but without the seed layer the easy axis of the Co layer in-plane [16]. In our film on hBN, the room temperature nucleation field, at which the magnetization reversal starts, is similar to those reported in conventional substrates [16,17]. This result indicates that the mechanism of perpendicular anisotropy in our film on the hBN is also from MgO/Co and Co/Pt interface anisotropy although the interface between Pt to hBN is a weak vdW interaction.

Since atomically thin hBN layer can serve as an excellent tunneling barrier [12], the MgO/Co/Pt/hBN system can also be employed for fabricating spin injection into 2D vdW materials by spin-polarized carrier tunneling. In this configuration, the thickness of the Pt layer is a parameter to determine the spin injection efficiency. While Pt is non-magnetic, the Pt layer in our multilayer system is magnetically proximitized, exhibiting magnetic property. In ref. [18], the characteristic thickness in Pt was found to be about 2.0 nm. We also need to control the coercive field of magnetic moment to realize the two distinguishable states of the parallel and anti-parallel configuration of ferromagnets [19-21]. Since the PMA is sensitive to the crystalline property of Pt and Co-O bonding strength, the thickness of Pt layer can serve as the control knob of coercive field. We fabricated MgO/Co/Pt films on hBN/SiO$_2$/Si where Pt layer thickness ranged 1-5 nm and measured Hall signal. We observed hysteresis both at the temperature of 4 K and at 300 K as shown in Fig. 2. We found that 3 nm Pt layer thickness exhibits the optimal performance with a sharp transition while below 1.0 nm did not show a well-defined hysteresis loop (not shown).

Conventional spintronics devices realized in 2D vdW materials has been utilized the ferromagnetic electrodes whose spin injection lies in-plane of the 2D channel [22]. The thickness dependent coercive field in MgO/Co/Pt/hBN structure can be employed for the PMA ferromagnet electrodes for injecting out-of-plane spin-polarized current into 2D vdW materials through the atomically thin hBN tunneling barrier. For this demonstration, we first use hBN encapsulated graphene device to realize injection and detection of out-of-plane spin in the 2D channel. The hBN/graphene/hBN hetero-structure was prepared by dry-transfer technique [23]. For the top layer hBN, we use monoatomic layer hBN for efficient tunneling of the carrier from the electrodes [12]. The right inset of Fig. 3(a) shows the topographic image of hBN(1ML)/Graphene/hBN(23nm) on $SiO_2$/Si substrate characterized by AFM. The measured surface roughness was ~60 pm, indicating atomically smooth surfaces and interfaces. Here, the bottom hBN layer serves as an atomically flat substrate to avoid the scattering caused by $SiO_2$ layer in the graphene channel [24]. On the top of this heterostructure, we fabricated the PMA electrodes by sputtering of Pt(4.0nm), Co(1.0nm) and MgO(1.5nm) using 400 nm thick PMMA for electron beam lithography resist for the lift-off process. Then, additional Au electrodes were fabricated to connect the PMA with the bonding pads. Ar-ion milling was employed to clean the surface of the PMA electrodes before depositing Au. Finally, the entire devices were gently annealed in vacuum at 350 °C. The left inset of Fig.3 (a) shows the optical microscope image of the final device. The pair of PMA electrodes is separated by 1.4 μm where each electrode has junction areas 1.2 (0.80 μm wide) and 0.52 $μm^2$ (0.60 μm wide), respectively. We designed the width and thickness of electrodes for different coercive fields, to detect total voltage from spin accumulation in the graphene channel.

Figure 3(a) shows magnetoresistance (MR) through PMA/hBN(1ML)/Graphene tunneling junction. As we sweep the out-of-plane magnetic fields, the observed MR exhibits two distinctive peaks around the field at B = 200 mT, indicating the spin accumulation in the 2D graphene channel can be detected by controlling the magnetizations of the PMA electrodes parallel and antiparallel [25-27]. A similar tunneling MR variation of two different ferromagnetic electrodes have been observed in graphene devices, but only in the in-plane magnetic field configuration [22]. Since our tunneling probes follow a local measurement scheme, the signal can include the MR of the PMA electrodes. However, Co/Pt layers tend to exhibit low MR ~0.02% [28], which is much smaller than the observed MR changes (~ 5%), and thus the observed MR peak is unlikely from the PMA electrodes themselves.

The PMA electrode combined with vdW heterostructure can be useful for spintronics applications employing TMD semiconductor. In ML TMDs, strong spin-orbit interaction splits the conduction and valence bands at the hexagonal Brillouin zone corner (termed as *K* and *K'* valley), coupling up (down) spin to *K* (*K'*) valley [2]. Therefore injecting spin-polarized carrier through a PMA electrodes can accumulate spin and valley in the channel. We chose ML $WSe_2$ as a p-type semiconducting channel that coupled with the PMA we developed. $WSe_2$ is a p-type semiconductor [29], where vdW junctions with Pt electrodes have been demonstrated to make Ohmic contacts [30]. Fig. 3(b) inset shows our device scheme for spin injection field effect transistor (FET) based on ML TMD channel. vdW stack of hBN(1ML)/$WSe_2$(1ML)/hBN(20nm) was prepared following a similar method described above. PMA [MgO(1.2nm)/Co(1.0nm)/Pt(2.0nm)] contact and reference electrodes made of Pt (20nm) and Co(20nm) were also fabricated by the sputtering method using a PMMA resist for the lift-off process.

Figure 3(b) shows the source-drain characteristics employing Pt, PMA, and Co as the source electrode and another Pt electrode as the common drain electrode. The current-bias characteristics show tunneling transport behavior with Pt and PMA electrodes at $T$ = 1.6 K, while the Co electrode exhibits significant suppression of the tunneling current. This result indicates that our PMA electrodes exhibit a comparable contact efficiency compared to that of the Pt electrodes, which is known to be highly efficient electrodes for $WSe_2$ [30]. We obtained ~> 10 nA current (at bias voltage -0.5 V) across the PMA/hBN/$WSe_2$ junction, which is almost an order of magnitude higher than that of the previously reported [Co(0.45nm)/Pd(1.5nm)]$_{20}$ PMA fabricated on $Al_2O_3$/$WS_2$(1ML) [31]. Fig. 3(c) also shows the device characteristic using the PMA-Pt electrodes pair as a function of back gate voltage. The increased drain current at negative gate voltages indicate the device is a p-type FET. We observed that $|V_{sd}|$> 0.2 V for a fulling functional FET with more than nA of on-current. While this value is still much better than those reported in the previously ($Ni_{81}Fe_{19}$) ferromagnetic contacts on $WSe_2$ [32], further improvement to realize low barrier contact is required to detect chemical potential changes due to the spin population changes in the channel.

In summary, we fabricate MgO/Co/Pt PMA electrode on hBN to utilize perpendicular anisotropy both from MgO/Co interface and Co/Pt interface. After vacuum annealing at $T$ = 350 °C, we find the perpendicular easy axis sustained in a wide temperature range 4-300 K with the thickness Pt layer down to 1.5 nm. We demonstrated that MgO/Co/Pt electrodes can be combined with van der Waals heterostructure consist of graphene and monolayer of $WSe_2$, where spin accumulation through the PMA charge injection and tunneling contact are realized. Our results open a new route towards studying spin dependent transport with perpendicular spin axis in vdW heterostructures.

**Acknowledgments**

This research was primarily supported by NSF (EFRI 2-DARE 1542807). H. I. acknowledges JSPS Postdoctoral Fellowship for Research Abroad (Grant Number 20140678). This work was performed in part at the Center for Nanoscale Systems (CNS), a member of the National Nanotechnology Infrastructure Network (NNIN), which is supported by the National Science Foundation under NSF award no. ECS-0335765. CNS is part of Harvard University.

**Figures**

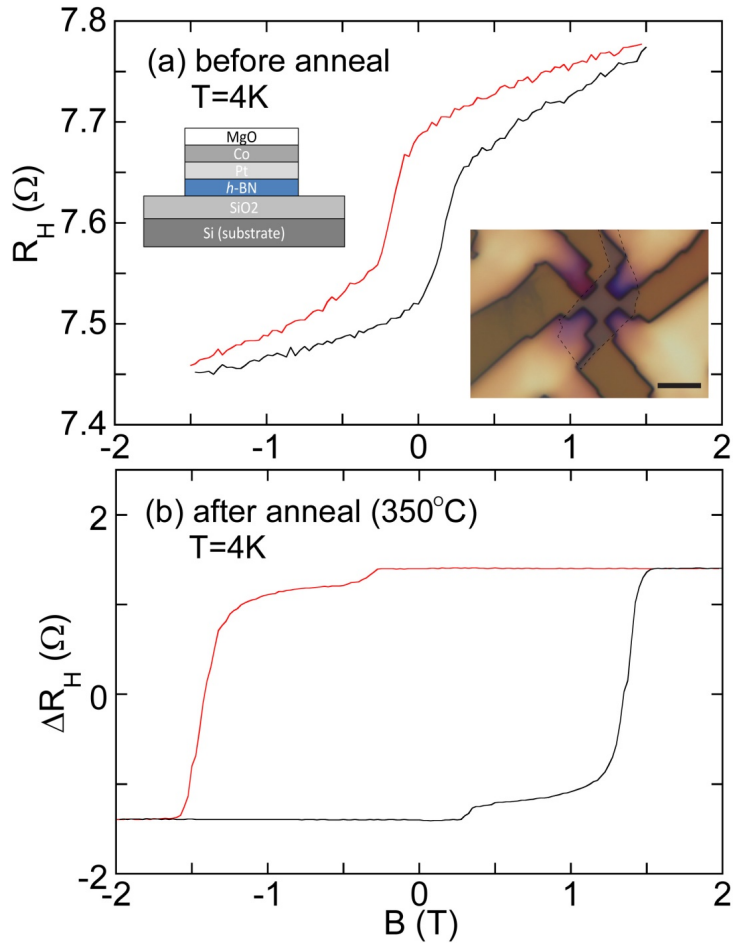

**FIG. 1.** (a) Hall signal of MgO(1.5nm)/Co(1.0nm)/Pt(5.0nm) film at the temperature of 4 K on hBN flake before annealing. The magnetic field was applied to perpendicular to the substrate. The film was etched into rectangular shaped channel connected to the electrodes. Insets show the cross-sectional schematic of the device (upper) and the optical microscope image of the device (lower). The scale bar represents 5 μm. The position of the hBN flake is marked d by the dashed line. (b) Hall signal at $T$ = 4 K after annealing at 350 °C under vacuum. Perpendicular anisotropy is drastically increased.

Fig.1  Idzuchi *et al*.

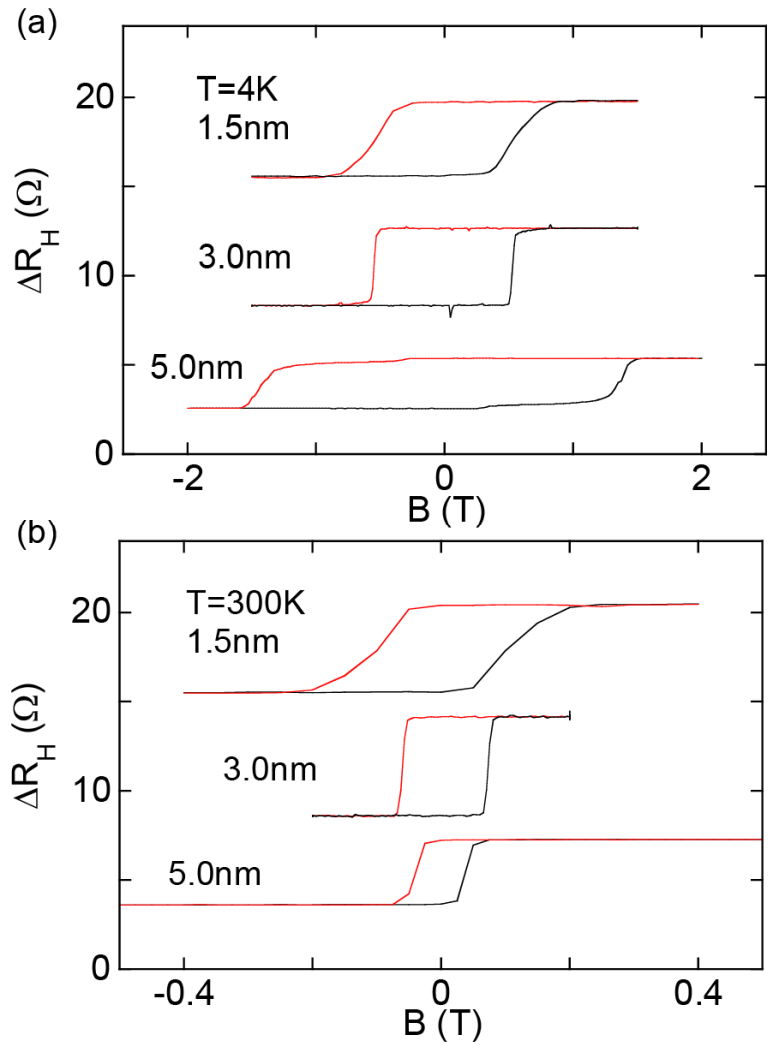

**FIG. 2.** (a,b) Hall signal of MgO/Co(1.0nm)/Pt(*x* nm)/ hBN flake with Pt thickness of *x*=1.5 nm, 3.0 nm, and 5.0 nm. Data taken at *T* =4 K and 300 K, exhibit perpendicular magnetic anisotropy.

Fig.2 Idzuchi *et al*.

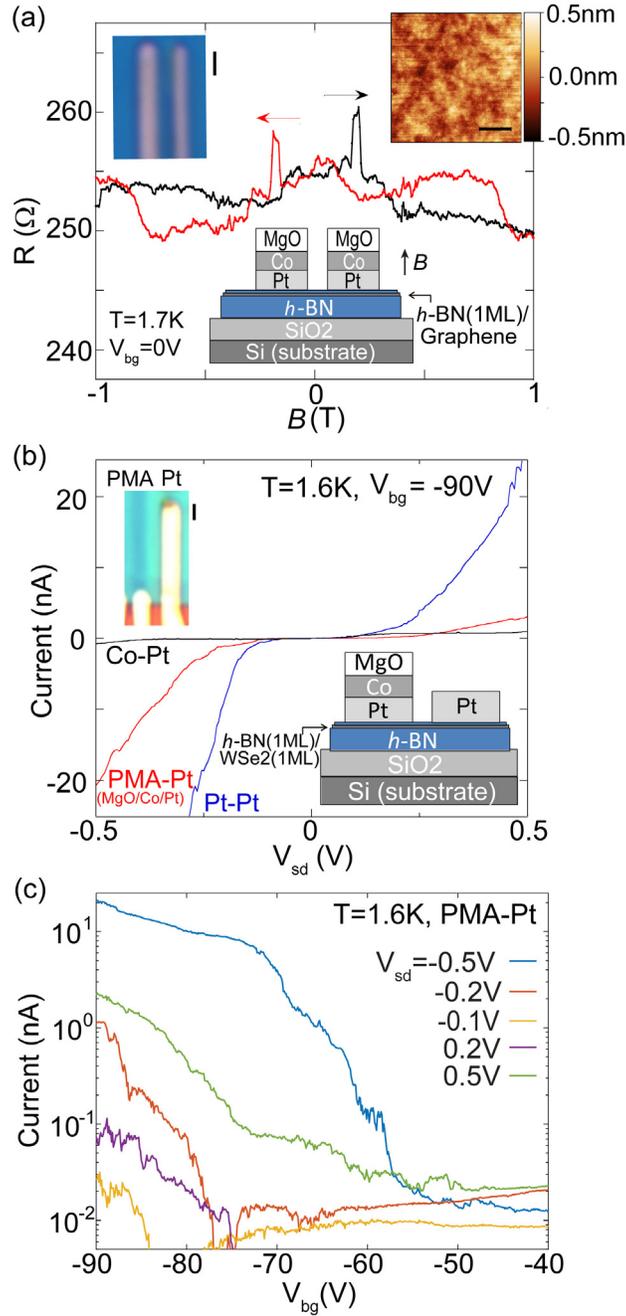

**FIG. 3.** (a) Tunneling device with MgO(1.5nm)/Co(1.0nm)/Pt(4.0nm)/hBN(1ML)/Graphene/hBN(23nm). The left inset shows the device image with a scale bar of 1 μm. The right inset shows AFM topography image of top hBN after the dry transfer process of all vdW layers. Resistance was measured at $T$ = 1.7 K with the applied magnetic field perpendicular to the substrate. The silicon back-gate was set at 0 V. (b) The source-drain current $I_{sd}$ as a function of the source-drain bias $V_{sd}$ in tunneling device with metal/hBN(1ML)/WSe$_2$(1ML)/h-BN(20nm) where metals are MgO(1.2nm)/Co(1.0nm)/Pt(2.0nm), Pt(20nm), Co(20nm), and Pt (20 nm). Monolayer WSe$_2$ channel is carrier-doped by the back-gate voltage of -90 V and measured at $T$ = 1.6 K. Left inset shows optical microscope image of the device with the scale bar representing 1 µm. (c) Field effect transistor behavior of 1ML WSe$_2$ with PMA at several fixed $V_{sd}$. $I_{sd}$ is shown as a function of the back-gate voltage between the PMA-Pt electrode pair as shown in (b).

Fig.3   Idzuchi *et al.*